# Theoretical p Mode Oscillation Frequencies for the Rapidly Rotating δ Scuti Star α Ophiuchi

Robert G. Deupree

Institute for Computational Astrophysics and Department of Astronomy and Physics
Saint Mary's University, Halifax, NS  B3H 3C3, Canada; bdeupree@ap.smu.ca

ABSTRACT

A rotating, two dimensional stellar model is evolved to match the approximate conditions of α Oph. Both axisymmetric and nonaxisymmetric oscillation frequencies are computed for 2D rotating models which approximate the properties of α Oph. These computed frequencies are compared to the observed frequencies. Oscillation calculations are made assuming the eigenfunction can be fit with six Legendre polynomials, but comparison calculations with eight Legendre polynomials show the frequencies agree to within about 0.26% on average. The surface horizontal shape of the eigenfunctions for the two sets of assumed number of Legendre polynomials agrees less well, but all calculations show significant departures from that of a single Legendre polynomial. It is still possible to determine the large separation, although the small separation is more complicated to estimate. With the addition of the nonaxisymmetric modes with |m| ≤ 4, the frequency space becomes sufficiently dense that it is difficult to comment on the adequacy of the fit of the computed to the observed frequencies. While the nonaxisymmetric frequency mode splitting is no longer uniform, the frequency difference between the frequencies for positive and negative values of the same m remains 2m times the rotation rate.

1. INTRODUCTION

Significant progress has been made over the last century in understanding the internal structure of stars and how stars evolve. Much of this knowledge is directly based on spherical stars, however, and there is a substantial gap in knowledge between spherical stars and stars which have significant nonspherical component. Included in the latter are stars which rotate appreciably. The difficulties are theoretical, computational, and observational, the last arising from relating observed quantities to physically relevant properties of the stars themselves. Before substantial progress can be made comparing models of rotating stars to observations, something must be done to restrict the otherwise large degrees of freedom, including the latitudinal variation of the surface properties, the inclination between the rotation axis and the observer, and the internal angular momentum distribution. It has been recognized that asteroseismology plus optical interferometry might be the combination that allows this to be done (Cunha, et al. 2007). An early attempt to determine the shape of the rapidly rotating star Achenar (Domiciano de Souza, et al. 2003) was clouded by the possibility that a circumstellar envelope was contributing to the oblateness measured by the interferometry as well as the stellar surface (Vinicius, et al., 2006; Kanaan, et al. 2008; Carcofi, et al. 2008; Kervella, et al. 2009), but it did lead to some attempts which successfully reproduced the observed shape (Jackson, MacGregor, & Skumanich 2004).

Both the structure and asteroseismology of rapidly rotating stars present challenges from a theoretical point of view. The calculation of multi-dimensional information for rotating models with reasonable physical input is possible (e.g., Clement 1978, 1979, Deupree 1995, Jackson, MacGregor & Skumanich 2005, Espinosa Lara 2010, Deupree 2011), although the numerical simulation of the redistribution of composition and angular momentum inside a rotating model during evolution is still incomplete (e.g., Tassoul & Tassoul 1982, 1995; Espinosa Lara & Rieutord 2007; Rieutord & Espinosa Lara 2009). Even with adequate models of rotating stars, challenges remain in the computation of oscillation mode frequencies. From an interpretation point of view, the fact that frequency splitting for nonaxisymmetric modes by rotation is nonuniform and exceeds the small and even the large separation at relatively low rotational velocities (e.g., Suarez, et al 2010, Deupree & Beslin 2010) adds complexity to mode identification. Also, the development of codes to compute oscillation frequencies which do not require a given mode to be defined by a single Legendre polynomial (e.g., Clement 1998, Lignières, et al. 2006, Reese, et al 2006, 2008, Lovekin, Deupree & Clement 2009) and their application with realistic models (e.g., Lovekin & Deupree 2008, Reese, et al. 2009) is a necessary step. While perhaps not on as well developed as one might like, the theoretical basis appears sufficiently satisfactory for an attempt when a suitable candidate becomes available.

Recently, α Oph (Rasalhague) has been observed interferometrically with the CHARA array (Zhao, et al. 2009) and asteroseismologically with the MOST satellite (Monnier et al. 2010). The star has a $V_{eq} \sin(i)$ of about 240 km s$^{-1}$, sufficiently large for the rotation to have appreciable effects on at least the surface conditions. The interferometry revealed that α Oph is seen nearly equator on, with a ratio of the polar radius to the equatorial radius of about 0.836. The observed luminosity and effective temperature (i.e., those quantities deduced assuming the observed flux is produced by a spherical star) range from 7880 to 8050 K for the effective temperature and from 25.1 to 25.6 $L_\odot$ for the luminosity (Blackwell & Lynas-Gray 1998, Malagnini & Morossi 1990, Monnier, et al. 2010). These values should be quite accurate given the facts that the star has a well determined parallax (see discussion in Gatewood 2005) and there is virtually no reddening. The combination of the deduced luminosity, effective temperature, and the observed oblateness, surface equatorial velocity, and inclination thus appreciably limit the possible models for this star. In particular, at least the surface rotation profile is less wide open than usual. These results place the star relatively close to the blue edge of the instability strip (e.g., Breger 2000, Xu et al. 2002). However, the perceived effective temperature and luminosity depend on the inclination between the observer and the rotation axis, and the actual luminosity and effective temperature for a rapidly rotating model seen equator on will both be higher than perceived, placing α Oph even closer to the blue edge. One desirable consequence of this is that atmospheric convection is not significant and can be ignored. To determine the precise relationship between the observed properties and the actual ones requires the spectral energy distributions from models of the surface effective temperature structure. The most rigorous approach would be to integrate the weighted intensity of the spectral energy distribution in the direction of the observer over all the visible surface using the latitudinal variation of the radius and local effective temperature to obtain the observed flux (e.g., Slettebak, Kuzma & Collins 1980; Linnell & Hubeny 1994, Frémat, et al. 2005; Gillich, et al. 2008) rather than specific assumptions being made about the limb and gravity darkening (e.g., Claret 2003, Reiners 2003, Townsend, Owacki & Howarth 2004, Monnier, et al. 2010). Rotating models whose properties match the observed spectral energy distribution, the surface equatorial velocity, and the oblateness would appreciably confine at least the surface stellar rotation properties. Until this is done one must be content with rotating models which match the approximate perceived effective temperature and luminosity along with the observed oblateness and surface equatorial velocity.

Analysis of the MOST data for α Oph revealed 57 oscillation frequencies which clearly must include both p and g modes. The combination of the comparatively large number of oscillation modes and the relatively detailed knowledge about the stellar properties make α Oph a good candidate to explore specific problems that might be encountered when trying to match the entire collection of data. To this end I have performed both 2D stellar evolution simulations and linear, adiabatic, nonradial oscillation calculations of α Oph. There are several objectives in this study. A major one is to learn what problems are associated with comparing observed and computed frequencies for moderately rapidly rotating stars for which the approximations made for small rotation are inadequate. A second objective is to determine how the oscillation frequencies are changed when the model parameters available, such as the rotation rate, mass, and composition profile are changed. In particular, the nature of scaling between models with only small differences in these parameters is examined. Supplemental to this is the determination of what, if any, useful role can be played by the often used asteroseismological parameters of the large separation and the rotational splitting. In particular, mode identification becomes a challenge for rotational velocities such as that of α Oph, and these and other tools may be helpful in this. Finally, some attempt is made to determine if the number of Legendre polynomials included in the computation of the linear, adiabatic eigenfrequencies is sufficient.

## 2. ROTATING STELLAR MODEL CALCULATIONS

The structure of rotating stellar models is computed with the 2D stellar evolution and hydrodynamics code, ROTORC (Deupree 1990, 1995). This code solves the conservation laws of mass, three components of momentum (there is an azimuthal momentum equation even though axial symmetry is imposed), energy, and hydrogen composition, as well as Poisson's equation. The independent variables are time, the fractional surface equatorial radius, and the colatitude. The surface equatorial radius is determined by requiring the integral of the density over the volume to be the desired mass. This integral, and the computation of the gravitational potential just outside the stellar surface as a boundary condition at each latitude in the 2D computational mesh, are included as part of the linearization in the Henyey technique. In these calculations equatorial symmetry is imposed.

To obtain the structure of a rotating model, one must specify the distribution of the composition and the distribution of the angular momentum. I evolved a 2.25 $M_\odot$ from the ZAMS through most of core hydrogen burning with local angular momentum conservation to determine the hydrogen composition profile. The ZAMS surface equatorial velocity was 281 km s$^{-1}$. A model with $X_c$ = 0.372, $M_{cc}$ = 0.28 $M_\odot$, L = 32.6 $L_\odot$, $T_{eff}$ = 8905K, and $V_{eq}$ = 221 km s$^{-1}$ was moderately close to the appropriate conditions and provides the base composition profile. Of course, there is no need for this profile to be the profile associated with α Oph, so the composition profile needs to be varied to examine its effects on the oscillation frequencies. The rotation profile of this model is neither conservative nor as smooth as one might like for computing oscillation frequencies. The latter is true because the material migrates through the non-Lagrangian mesh, and the numerically unstable centered advective terms are just barely stabilized to keep the numerical diffusion low but numerically accurate. For these reasons, a rotation distribution was imposed on this model and the structure reconverged. Uniform rotation with the same value of the surface equatorial velocity was assumed for simplicity and as a plausible place to start, although the structure was also computed for a modestly differentially rotating model.

I have computed a number of uniformly rotating models whose fundamental properties are listed in Table 1. All models have 581 radial zones and ten angular zones. The effective temperature given in Table 1 is computed by dividing the luminosity by the total surface area; the effective temperature and luminosity one would observe depends on the observer's inclination to the rotation axis and the pole to equator temperature variation in a reasonably well determined way (e.g., Collins 1966; Collins & Harrington 1966; Maeder & Peytremann 1970; Linnell & Hubeny 1994; Reiners 2003; Townsend, Owocki & Howarth 2004; Frémat, et al. 2005; Gillich, et al. 2008). The "V" sequence of models were obtained from the first one by imposing uniform rotation with a higher surface equatorial velocity while keeping the total mass fixed and the internal composition profile the same as a function of radial zone number. Because these 2D calculations are non-Lagrangian, the composition profile does change slightly as a function of interior mass from one model to the next. Increasing the rotation rate leads to slightly higher central temperatures, and the composition being uniform a few hundredths of a pressure scale height beyond the formal boundary of the convective core for the new model. These effects, particularly the implicit introduction of convective core overshooting which increases the luminosity on the main sequence (e.g., Maeder & Mermillod 1981; Stothers & Chin 1981; Doom 1985; Bertelli, Bressan, & Chiosi 1985; Maeder & Meynet 1989), allow the luminosity to increase as the rotation rate increases instead of decreasing with increasing rotation as it does on the ZAMS. Model M1 is obtained from model V245 by decreasing the total mass and then changing the surface equatorial velocity until the surface shape of M1 is the same as that of V245. The surface equatorial velocity required to do this is 241.5 km s$^{-1}$. I have found (Deupree 2011) that keeping the surface shape the same when changing other parameters allows many ratios, such as the effective temperature ratios at any two latitudes, to remain the same. Model C1 was obtained from model V225 by moving the composition profile inward one radial zone and then adjusting the surface equatorial velocity until the surface shape for the two models is the same. Model C2 was similarly produced from model V245 by a two radial zone shift inward. The surface equatorial velocities required were 227.5 km s$^{-1}$ for model C1 and 251 km s$^{-1}$ for model C2, respectively. Model D1 was computed from model V230 by imposing differential rotation with an algorithm generalized from one used by Jackson, MacGregor, & Skumanich (2004)

$$\Omega = \frac{\Omega_0}{1 + \alpha \varpi^\beta}$$

$\Omega_0$, α, and β are constants, and $\varpi$ is the distance from the polar axis in units of the surface equatorial radius. Note that this equation has the rotation rate decreasing with increasing distance from the rotation axis. The variable used in ROTORC is the rotation velocity, which is obtained from the above equation to be

$$V_\phi(r,\theta) = \frac{V_{eq}\varpi(1+\alpha)}{1+\alpha\varpi^\beta}$$

Model D1 has α = 2 and β = 0.2. Interestingly, Table 1 shows that both the equatorial radius and the polar radius for models D1 and V240 are very close to the same. It should be noted that there are differences in the radii for the two models at mid latitudes so the shapes are not quite the same. Nevertheless, it will be interesting to see how the oscillation frequencies of the two models relate.

The "V" sequence was computed to obtain rotating models with the approximate α Oph v sin *i* and oblateness. Models V235, V240, and V245 all match these two parameters to within quoted errors (Monnier, et al. 2010). Models M1, C2, and D1 test the effects of varying the mass and composition profile and the effect of differential rotation, respectively, while still matching these two properties. The equatorial effective temperatures, expected to be lower than the deduced effective temperatures for a star seen equator on, range from 7949K to 8111K for this set of models except for model M1, which has an effective temperature of 7757K.

## 3. AXISYMMETRIC MODE OSCILLATIONS

I have computed the axisymmetric mode oscillation frequencies for all pulsation modes with frequencies between 15 and 50 cycles per day using the linear, adiabatic, nonradial oscillation code of Clement (1998), as modified by Lovekin, Deupree, & Clement (2009). Higher frequencies were computed for some models as needed. This code is suitable for rapidly rotating stars because it allows each mode to be represented by the sum of several spherical harmonics rather than just one. While this is a requirement for rapidly rotating models, it does make mode identification more complex because the usual latitudinal mode identifier, ℓ, is no longer a valid quantum number. Increasing the number of spherical harmonics, or "basis functions", increases the accuracy of the frequencies, but it also adds many more modes that must be computed and makes the identification of specific modes harder. Here I shall use six basis functions most often, but will use eight basis functions to test how reliable the frequencies are.

Axisymmetric modes in most of this frequency range are expected to be p modes, although some of the observed modes could be either avoided crossing g modes or nonaxisymmetric g modes shifted into at least the low frequency part of this region if -m, which remains a valid quantum number, is sufficiently high. It should be pointed out that the usefulness of the radial quantum number, n, is significantly degraded through unknown amounts of avoided crossings and through there being different numbers of nodes in the equatorial and polar directions in these rapidly rotating models (e.g., Clement 1998).

The relationship between the frequencies computed for the different models is examined in Table 2. This table may be understood in the following way. The same pulsation mode is identified for two models, listed as A and B in the table. The ratio between the frequencies for the same mode in the two models is computed. This is repeated for about 30 other modes. The frequency ratio, averaged over all these modes, is presented as $\omega(A) / \omega(B)$ in the table. The standard deviation of this ratio from the average is also given. Table 2 shows that the frequencies do scale to some extent from one model to another, at least at the higher frequencies where there is less ambiguity about mode identification. The designation "V240(8)" and "V245(8)" refer to computing the frequency ratio for these two models using eight basis functions instead of six. Table 2 shows that there is some difference in the ratio produced by including more basis functions but at this point it does not overwhelm the effects we are examining. The existence of scaling between two models, even if they cannot differ in properties very much, is important because it would allow one to change the properties of a model which fairly closely matches observed frequencies to obtain a model which would match the observed frequencies even better.

Two modes being the "same" perhaps require some definition. The individual frequencies used for computing the frequency ratio between two models were selected based both on the similar nature of the surface horizontal shape of the radial perturbation eigenfunction between the two frequencies and on the frequencies being members of a set in which the surface horizontal shape changes in the same way from one frequency to the next in the two models. The differences between the individual models used for the ratios in Table 2 are sufficiently small that this comparison is relatively easy, and it did not seem to matter whether we used six of eight basis functions. It becomes progressively more difficult as the differences between the properties of the two models increase.

This table presents several interesting features. The frequencies decrease by about two to three per cent for every 5 km s$^{-1}$ increase in the surface equatorial velocity at these velocities with a standard deviation about ten per cent of the decrease. Note that the standard deviation is about five times smaller where the two models in a comparison have the same surface shape, indicating that scaling is not as exact when the shape of the surface changes. This is possibly another reason for using the surface shape to indicate the degree of rotation. For the comparison of models with the same surface shape, it is reasonable to ask if they follow the period – root mean density scaling law because the surface radii at all latitudes change by the same scale factor. The frequency ratio deduced from this scaling law when the mass is changed is 1.0098, somewhat higher than the 1.0091 frequency ratio found. The difference is outside the standard deviation, but there is some possible variation produced by the fact that the surface radius at each latitude is quantized by the radial zoning of the model. Similarly, the period – root mean density relation gives frequency ratios of 1.0336 and 1.0806 for the two cases where the composition changes, to be compared with the computed frequency ratios of 1.0339 and 1.0796, respectively.

It is interesting to compare the differentially rotating model not with the model from which it was derived and with which it shares the same surface equatorial velocity (V230), but with the model which has nearly the same equatorial and polar radius (V240). The frequencies for models D1 and V240 are nearly the same, suggesting it would be exceedingly difficult to distinguish between a slightly differentially rotating model and a model uniformly rotating at a slightly higher surface equatorial velocity. It may be that the nonaxisymmetric modes resolve the issue, although those modes present challenges of their own. This frequency ratio result is also consistent with the result of Lovekin, Deupree, & Clement (2009) that differential rotation laws of this form essentially produce frequencies similar to those of a slightly more rapid uniformly rotating model.

Another frequency comparison among the different models is the large separation. The determination of the large separation requires modes with higher values of n than are present in all but two or three of the observed modes in α Oph. Nevertheless, the large separation can still be useful in identifying the sequence of modes that corresponds to some particular value of ℓ in a nonrotating model. This does not imply that one knows what the value of ℓ is, merely that one can identify the particular members that have the property of having the same effective ℓ. This identification becomes more difficult at lower frequencies because the evolution of the shape of the eigenfunction appears to change more quickly with n at lower n. However, knowing the value of the large separation does allow the determination of at least some member of the same sequence at lower n, and it is thus useful to determine the value of the large separation. Thus, I have identified modes over some frequency interval whose horizontal surface shape seems to evolve in a sufficiently slow and recognizable way to allow computation of the large separation. These results are presented in Table 3, along with the standard deviation over about 15 frequency pairs for each model. Generally speaking, the large separation decreases as the rotation rate (and the stellar volume) increases. Similarly, shifting the composition profile inward decreases the model volume and makes the large separation larger. Decreasing the mass also decreases the volume, but the net effect is to decrease the frequencies as well because the mass enters into the scaling relation. Making the model rotate differentially decreases the large separation from that of the uniformly rotating model from which it was created (V230), but the large separation is very close to that of the model which has nearly the same surface equatorial and polar radii (V240). This agrees with the result of Lovekin, Deupree, & Clement (2009) that differential rotation does not have an appreciable effect on the large separation except at extreme differential rotation.

An echelle diagram of the higher frequency modes is shown for model V240 in Figure 1. In this figure I have labeled six curves with a designation "A" to "F", representing the six latitudinal mode configurations for the six basis functions. Note that the modes between two sequences occasionally get close together, at which point each of the eigenfunctions of the two modes appear to have attributes of both of the sequences as determined from the frequencies above and below those which are close together. The surface horizontal eigenfunction properties of sequences C and D do appear to cross where the frequencies of the two sequences become very close, but the properties of both sequences A and B appear to change going through the frequencies where the two sequences come close together, and neither looks quite like it did at lower frequencies than that at which the frequencies became close. Without prejudicing the issue of whether these represent true crossings or avoided crossings, these will be referred to as "interactions".

The individual latitudinal mode configurations show enough curvature that only modest estimates of the small separation can be made. Even these are only general because we have no indication how the different latitudinal configurations relate to each other. For example, it is not obvious that there is any way to determine which configurations would have the equivalent of $\Delta\ell = 2$ (or even if such a distinction makes any sense) without computing a full sequence of models and modes from slow rotation to the current rotation rate. About the only concrete statement that can be made is that, with certain exceptions, the small separation is consistent with the average value of about 0.63 cycles per day.

I have computed frequencies with eight basis functions for models V240 and V245. An echelle diagram for these modes of model V240 is presented in Figure 2. Of course there are two more latitudinal mode configurations. The mode configurations are labeled "A" through "H", with the relationship between the same alphabetic designation being determined by the matching of the highest frequencies. The large separations differ by a little more than one per cent. There are several more mode interactions in the larger basis function data. Neglecting the division into $\ell$ series and just comparing the closest frequencies, the mean difference between the frequencies with six and eight basis functions is about 0.11 cycles per day. The largest difference between the two frequency sets is about four times larger than the mean, and it occurs where one frequency set shows an interaction and the other does not. In fractional terms the mean difference is about 0.26%, with the highest difference slightly less than 0.9%. Comparing the frequencies of a specific $\ell$ series for the two sets of basis functions, the largest difference is about 1.7% and occurs at low frequency. Above 52 cycles per day all differences are less than 0.6%. This suggests that the association of frequencies with $\ell$ series may not be entirely straightforward at lower frequencies. Clearly, the calculation of the frequencies for models rotating this rapidly has nothing like the accuracy of the observations themselves.

While the frequencies computed with the two numbers of basis functions are within a per cent of each other, such agreement cannot be expected between the eigenfunctions themselves. After all, the fact that fairly marginal eigenfunctions give fairly decent frequencies is what made the variational technique (e. g., Chandrasekhar and Lebovitz 1968) useful. Here I will focus on the horizontal variation of the radial component of the displacement at the model surface because it should be useful in determining what modes can be observed if the inclination is known, as it is here. All of these results are for model V240. A comparison of this component for six and eight basis functions as a function of the colatitude is presented in Figure 3. Recall that the scaling is arbitrary in these linear calculations. These modes are from the $\ell = C$ set. From the eight basis function plot it is clear that this component of the eigenfunction is appreciably larger near the polar axis than elsewhere. Clearly, the higher order Legendre polynomials are required to produce the approximately constant, low amplitude seen at mid latitudes and, to a lesser extent, near the equator. The two frequencies agree to within 0.005 per cent.

It could be argued that this is not the best case because the coefficient of $P_{10}$ in the six basis function case is relatively large compared to the other coefficients. However, this does not appear to be a significant factor. In Figure 4 I compare the same component of the eigenfunction for an $\ell$ = A mode in which the two largest coefficients in both the six and eight basis function cases are for Legendre polynomials of order eight or less. It is clear that this eigenfunction does not appear to be any closer to be determined than the one in Figure 3. While this many basis functions may be close to providing reasonably reliable frequencies, more basis functions are required to obtain (at least this part of) the eigenfunction.

With these estimates of the dependence of the frequency variation with the number of basis functions providing some estimate of their reliability, I wish to examine how the computed modes compare to the observed mode frequencies. The object at this stage is to see if the axisymmetric modes of any one model provide a better match to the observed data than those of the other models. I included models V225, V230, V235, V240, and V245. The answer is effectively negative. Model V225 was substantially less successful, but all the others matched about fifteen of the 35 observed modes to within the 0.11 cycles per day mean difference between the six and eight basis function calculations. It is also true that the agreement was not significantly changed for either models V240 or V245 if I used the eight basis function calculations instead of the six basis function calculations. This cannot be considered great agreement, but then there is no reason to believe that all of the observed modes are axisymmetric. I shall now turn to the calculation of nonaxisymmetric modes for one of these models.

## 4. NONAXISYMMETRIC MODE OSCILLATIONS

Rapid rotation introduces complexities into nonaxisymmetric modes as well. One complication of rapid rotation is that the frequency spacing between adjacent members of a given multiplet becomes nonuniform (e.g., Espinosa, Pérez Hernández & Roca Cortés 2004; Reese, Lignières & Rieutord 2006). Furthermore, as the rotation increases the frequency difference between adjacent m values will become larger than the frequency difference first between adjacent $\ell$ values and then between adjacent n values (e.g., Suárez, et al. 2010; Deupree & Beslin 2010). One can expect both issues to significantly affect the frequency spectrum of models rotating as rapidly as α Oph.

I have computed the oscillation frequencies for model V230 assuming six basis functions and latitudinal modes symmetric about the equator for |m| ≤ 4. The echelle diagram for this collection, plus the axisymmetric mode frequencies and the observed frequencies are shown in Figure 5. The observed modes have been artificially offset vertically to identify them. Clearly, the large number of modes decreases the "open space" in frequency so that most observed modes have modes with very similar computed frequencies. Perhaps a more useful way to examine this is Figure 6, which shows, as a function of the observed frequency, the magnitude of the frequency difference between the observed frequency and the closest computed frequency, regardless of any property the computed frequency may have. About 19 of the 35 observed modes in this frequency window are matched within the observational error, and another nine within twice the observational error. Four of the remaining seven are more than three times the observational error, but all the computed modes can still have their frequencies shifted by amounts large compared to this by adding more basis functions. These results do not mean that this is a good model, but rather that more information will be required to eliminate or constrain many of these computed modes before meaningful comparisons between observations and theory can be made.

I reexamined the issue of patterns in the nonaxisymmetric mode frequency spectrum. It turns out that, while the frequency difference between adjacent members of a multiplet is not uniform, it is constant between specific members of the multiplet. I find that the quantity

$$D_m \equiv \frac{\omega_{-m} - \omega_m}{2m}$$

is very nearly constant as a function of both n and m. Table 4 lists the value of $D_m$ and its standard deviation, both converted to cycles per day, based on twenty to thirty values of $D_m$ for each m. The values are close to the rotation frequency for this model of 1.636 cycles per day. This result prompted me to reexamine the modes computed by Deupree and Beslin (2010), and they show that $D_m$ is approximately constant there as well.

The constancy of $D_m$ does not mean that the members of the multiplet are uniformly spaced. This can be seen in Table 5, for which I present the average frequency, $\omega_{am}$, again converted to cycles per day, given by

$$\omega_{am} = \frac{\omega_{-m} + \omega_m}{2}$$

for a specific multiplet as a function of m. For m = 0, I present the three computed modes which cover the range of $\omega_{am}$ for the various values of m. There are two possible mode pairs for m = 4, and the average frequency is included for both. If all the modes in the multiplet were uniformly spaced, $\omega_{am}$ would be the same for all m. The fact that $\omega_{am}$ is not constant means that the modes for different m's are shifted with respect to each other, even though $D_m$ is effectively constant for all m. This does mean that a frequency very close to the rotation rate should appear in the Fourier transform of the frequency spectrum, although sophisticated methods may be required to extract it from this information.

It is worth examining whether there is any evidence of $D_m$ in the α Oph oscillation data. Taking the Fourier transform of the oscillation data yields no useful information, as might be expected from the results of Deupree and Beslin (2010). Looking at the 35 frequencies above 16 cycles per day individually, I find fourteen cases in which the frequency difference between two modes is a multiple of the rotation frequency of 1.65 cycles per day (Monnier, et al. 2010). However, five of these frequency differences are odd multiples of the rotation frequency, whereas the definition of $D_m$ requires the frequency difference to be an even integer multiple of the rotation rate. All nine of the allowed (i.e., even integer multiples) frequency differences yield a rotation rate of 1.6472 with less than one percent variation. Six of these correspond to m = 2, two to m = 1, and one to m = 4. I repeated the exercise assuming a rotation frequency of 2.518 and 1.259 cycles per day for a comparison. The higher frequency case yielded ten cases in which the frequency interval was a multiple of this assumed rotation rate, but five of these were odd integer multiples. The equivalence of the number of odd and even multiple cases suggests this looks like chance. However, the lower frequency choice produced fifteen cases with multiples of the assumed frequency, five of which were odd multiples, again with a total variation of one percent for the remaining ten. This suggests that knowledge of the rotation rate may help identify members of a multiplet, but that using the observed frequency separations for rapidly rotating stars to determine the rotation rate will require more sophisticated analysis.

## 5. DISCUSSION

I have computed the structure and oscillation frequencies for several 2D rotating stellar models which may approximate the properties of α Oph. Stars which rotate this rapidly require several spherical harmonics to calculate accurate oscillation frequencies; my comparison between six and eight spherical harmonics provides agreement in the frequencies to 0.25% on average and about 0.9% in the worst case, assuming that the closest frequencies between the two sets match the same mode. However, a comparison of the horizontal surface variation of the radial displacement reveals that, in most cases, the six spherical harmonic and eight spherical harmonic eigenfunctions do not resemble each other very closely.

A comparison between the 35 observed and computed frequencies for all $|m| \leq 4$ for frequencies above 15 cycles per day looks impressive at face value – the largest percentage difference in frequency between an observed mode and a computed mode is less than 0.3%. This becomes less impressive with the realization that there are 589 computed modes in the frequency interval, so that the frequency difference between a computed and observed mode for three cases is larger than the mean spacing between computed modes. For two observed frequencies, the computed mode with the closest frequency has $|m| = 4$. These are the only cases where $|m| = 4$ significantly improves the frequency difference, so it does not appear that limiting consideration to $\ell \leq 3$ and hence $|m| \leq 3$ notably improves the situation.

These modes have been used without regard to their surface properties, so that some modes could potentially be removed from consideration because their horizontal properties at the surface might make them unlikely to be seen via photometry. However, as we have seen, these surface properties are not as settled as the frequencies themselves, and it may take several more spherical harmonics in the sum before they settle down. Of course, each new spherical harmonic adds more frequencies to the computed frequency total count.

This comparison between observed and computed frequencies for a moderately rapidly rotating star indicates that there are several issues that must be resolved before one can comfortably conclude that any computed model of a rotating star is an adequate and constraining fit to an individual star if it is sufficiently rapidly rotating.

There are a few areas in which further research could refine the range of acceptable conditions. Probably the most important of these is computing the spectral energy distribution for these rotating models and comparing it with that observed. Such a comparison may place further constraints on the range of surface temperature conditions allowed and perhaps on the shape of the rotation profile as a function of latitude. Further work on the surface horizontal shape of the surface displacement may reduce the number of possible theoretical modes that could produce observable amplitudes. The problem of comparing observed and theoretical modes for rapidly rotating stars is complex, but it may become tractable if the stellar conditions can be suitably constrained.

Table 1

Properties of Models Considered

| Model ID | $V_{eq}$ (km s$^{-1}$) | Mass ($M_\odot$) | Luminosity ($L_\odot$) | $T_{eff}$ (K) | $R_{eq}$ ($R_\odot$) | $R_{pole}/R_{eq}$ | $T_{pole}/T_{eq}$ |
|---|---|---|---|---|---|---|---|
| V221 | 221. | 2.25 | 33.95 | 8776. | 2.669 | 0.8687 | 1.143 |
| V225 | 225. | 2.25 | 34.31 | 8747. | 2.707 | 0.8640 | 1.149 |
| V230 | 230. | 2.25 | 34.84 | 8708. | 2.763 | 0.8567 | 1.153 |
| V235 | 235. | 2.25 | 35.45 | 8668. | 2.825 | 0.8463 | 1.170 |
| V240 | 240. | 2.25 | 36.19 | 8624. | 2.897 | 0.8380 | 1.181 |
| V245 | 245. | 2.25 | 37.05 | 8579. | 2.977 | 0.8293 | 1.191 |
| M1 | 241.5 | 2.18 | 33.31 | 8371. | 2.965 | 0.8293 | 1.191 |
| C1 | 227.5 | 2.25 | 33.28 | 8778. | 2.648 | 0.8640 | 1.149 |
| C2 | 251. | 2.25 | 34.36 | 8639. | 2.827 | 0.8293 | 1.191 |
| D1 | 230. | 2.25 | 35.96 | 8606. | 2.891 | 0.8380 | 1.183 |

Table 2

Frequency Ratio between Models

| A | B | ω(A) / ω(B) | σ |
|---|---|---|---|
| V221 | V225 | 1.0181 | 0.0019 |
| V225 | V230 | 1.0247 | 0.0028 |
| V230 | V235 | 1.0232 | 0.0030 |
| V235 | V240 | 1.0305 | 0.0035 |
| V240 | V245 | 1.0310 | 0.0030 |
| V240(8) | V245(8) | 1.0336 | 0.0020 |
| V245 | M1 | 1.0091 | 0.00057 |
| C1 | V225 | 1.0339 | 0.00047 |
| C2 | V245 | 1.0796 | 0.00058 |
| D1 | V240 | 1.0019 | 0.0021 |

Table 3

Large Separation for Rotating Models

| Model | Large Separation (cycles / day) | Standard Deviation (cycles / day) |
|---|---|---|
| V221 | 4.1216 | 0.1499 |
| V225 | 4.1052 | 0.1612 |
| V230 | 4.1102 | 0.1637 |
| V235 | 3.9805 | 0.1599 |
| V240 | 3.8873 | 0.1448 |
| V240(8) | 3.9264 | 0.1373 |
| V245 | 3.8042 | 0.1020 |
| V245(8) | 3.8735 | 0.0617 |
| M1 | 3.7690 | 0.1373 |
| C1 | 4.1807 | 0.1990 |
| C2 | 4.0901 | 0.1335 |
| D1 | 3.8370 | 0.1650 |

Table 4

Nonaxisymmetric Mode Frequency Separation $D_m$ as a Function of m

| m | $D_m$ (cycles / day) | $\sigma$ (cycles / day) |
|---|---|---|
| 1 | 1.604 | 0.043 |
| 2 | 1.596 | 0.028 |
| 3 | 1.612 | 0.011 |
| 4 | 1.614 | 0.013 |

Table 5

Average Frequencies for a Given Multiplet

| m | $\omega_{am}$ (rad /day?) |
|---|---|
| 0 | 42.295, 42.997, 43.427 |
| 1 | 42.483 |
| 2 | 43.119 |
| 3 | 42.870 |
| 4 | 42.191, 43.441 |

Figure Captions

Fig. 1 – Echelle diagram for the six ℓ sequences for six basis function frequency oscillation calculations of model V240. A given sequence, identified by an alphabetic charater, is defined by the surface horizontal variation of the radial displacement. The approximately vertical nature of the curves means that the large separation is reasonably well defined. Sequences C and D do appear to cross in that the shape of the surface horizontal variation of the radial displacement follows the individual modes as shown.

Fig. 2 – Echelle diagram for the eight ℓ sequences for eight basis function frequency oscillation calculations of model V240. A given sequence is defined by the surface horizontal variation of the radial displacement. The large separation and the scale are the same as shown in Figure 1. The identification of sequences A through F is based on the commonality of frequencies with those same sequences in Figure 1. It should be noted that the surface horizontal variation of the radial displacement of a given sequence in Figure 1 does not much resemble that of the same sequence in this Figure.

Fig. 3 – Comparison of the surface horizontal variation of the radial displacement for one mode of ℓ sequence C for six basis functions (solid) and eight basis functions (dash). It is possible to see how the addition of higher order spherical harmonics affects the horizontal behavior of this mode whose radial motion is largely at the pole.

Fig. 4 – Comparison of the surface horizontal variation of the radial displacement for one mode of ℓ sequence A for six basis functions (solid) and eight basis functions (dash). Unlike the comparison in Figure 3, it is not clear in this case what the horizontal variation with a large number of spherical harmonics would be even though neither curve plotted is dominated by the highest spherical harmonic allowed.

Fig. 5 – Echelle diagram of the α Oph observed frequencies (circles) and the calculated frequencies for model V230 for $|m| \leq 4$. The computed frequency symbols are given by squares ($m = 0$), diamonds ($|m| = 1$), inverted triangles ($|m| = 2$), triangles ($|m| = 3$), and right facing triangles ($|m| = 4$). The observational data have been artificially offset vertically to highlight them. The profusion of computed modes produces so many matches to the observed frequencies that distinguishing between models may be difficult.

Fig. 6 – Comparison between the minimum frequency difference between an observed frequency and a computed frequency for the computed frequencies of model V230 ($|m| \leq 4$). The observed frequency uncertainty and twice the uncertainty quoted by Monnier, et al. (2010) are shown by the two horizontal dashed lines.

Figure 1

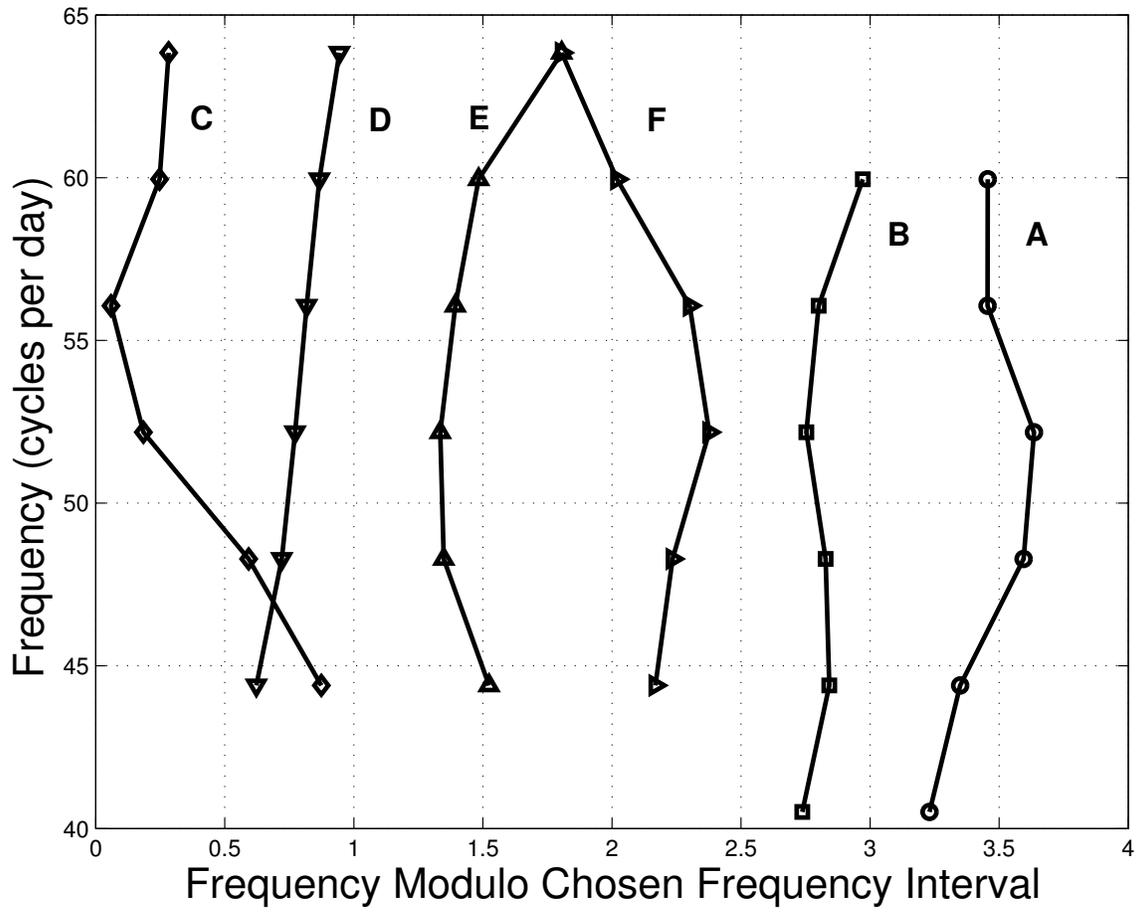

Figure 2

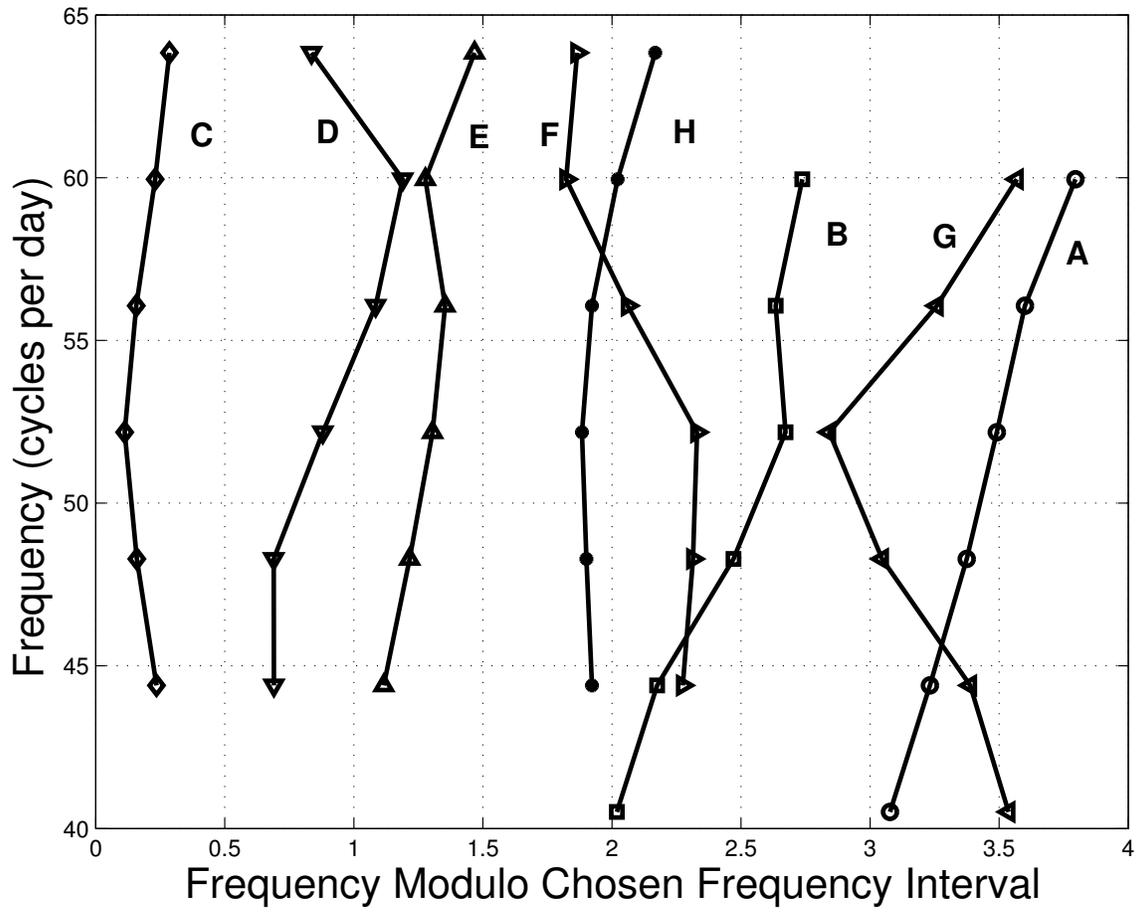

Figure 3

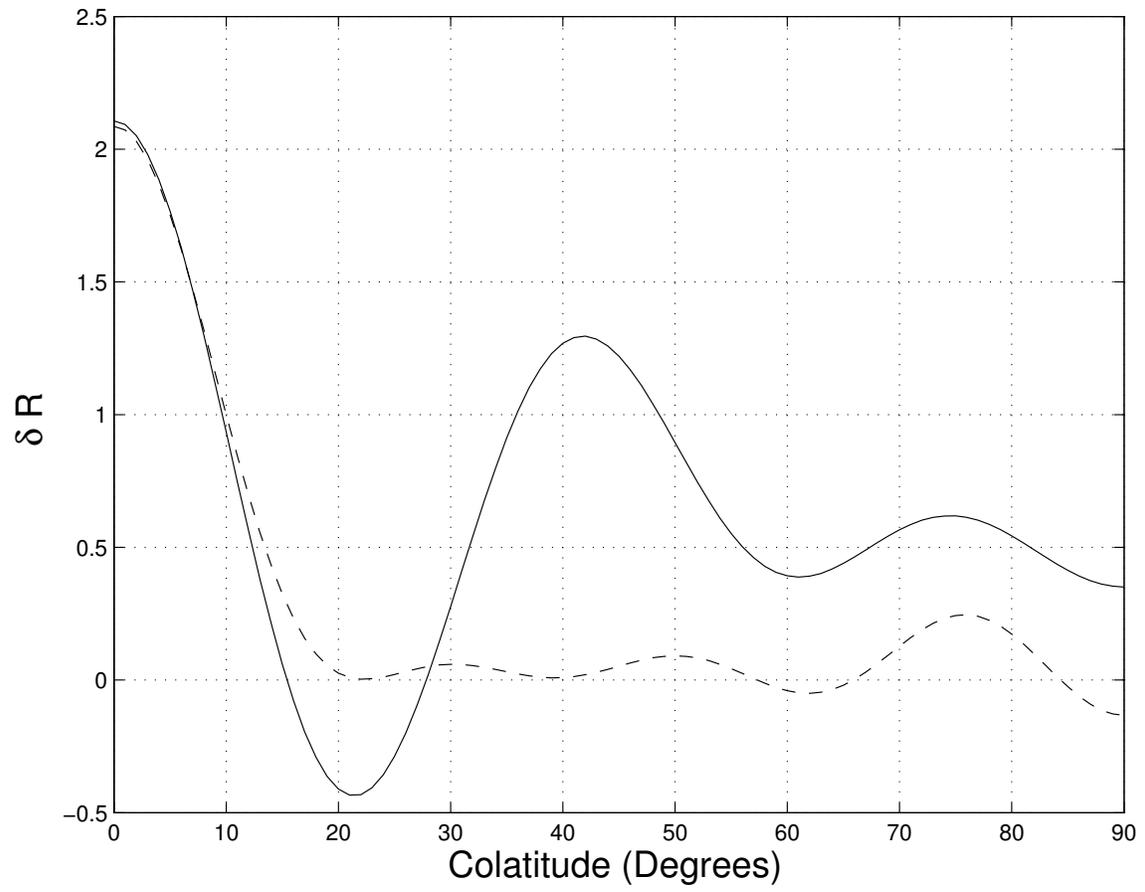

Figure 4

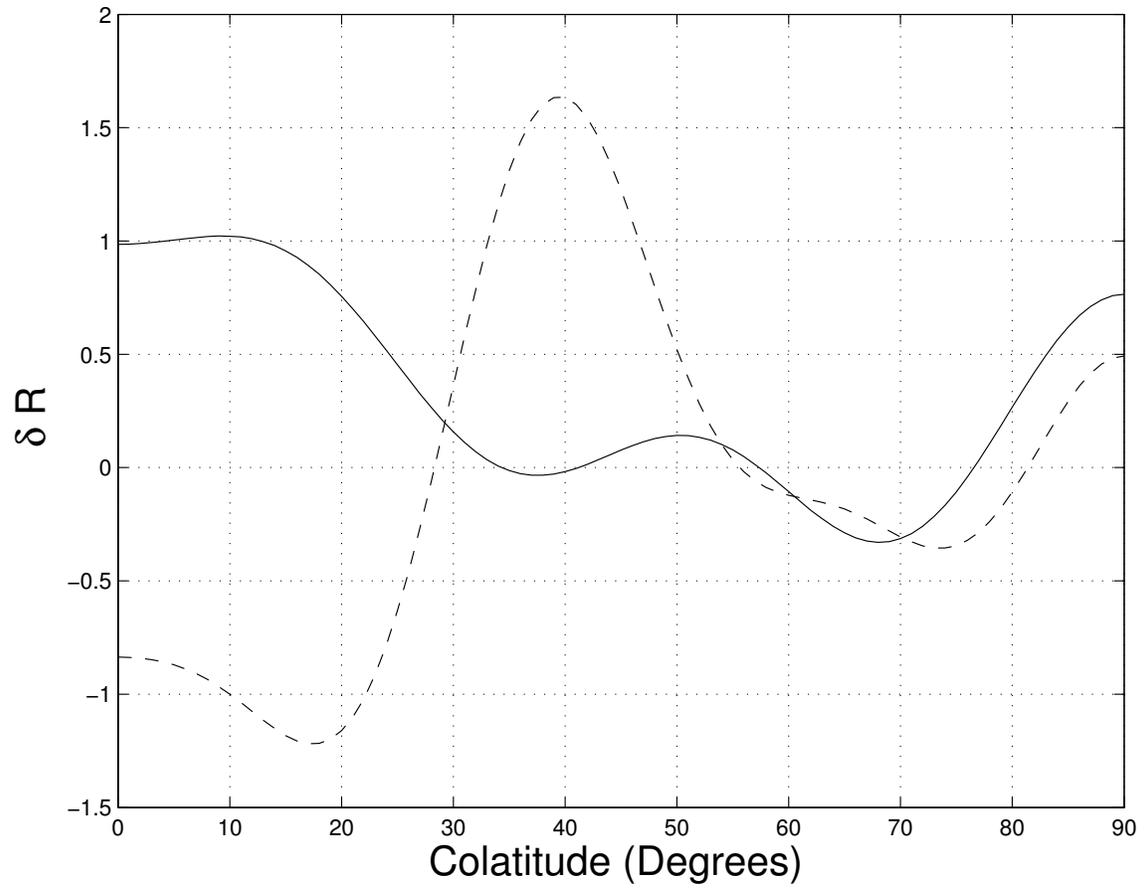

Figure 5

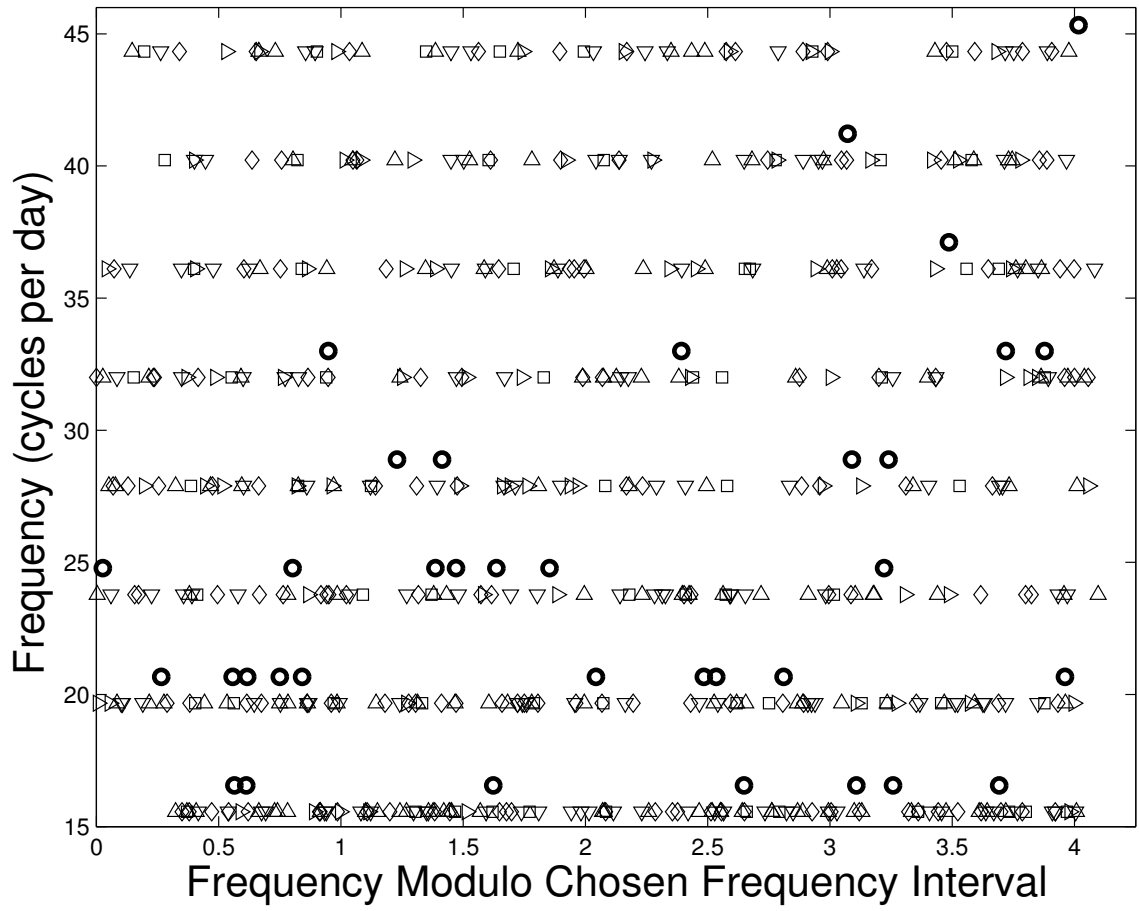

Figure 6

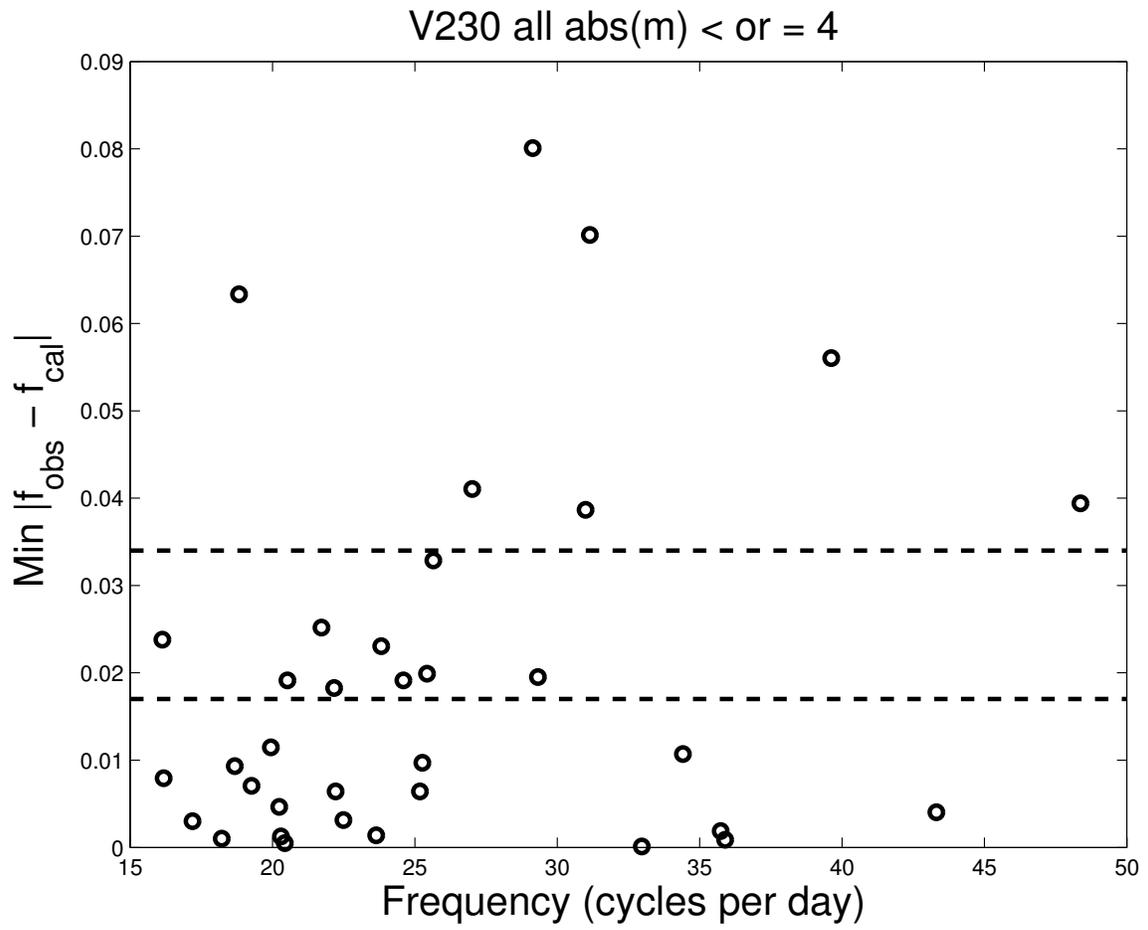